\documentclass[10pt,letterpaper,twocolumn]{article} 

\usepackage{ol2}
\usepackage[draft]{hyperref}
\usepackage{amsmath}

\begin{document}

\twocolumn[ 

\title{Quantum statistical signature of $\mathcal{PT}$ symmetry breaking}


\author{Stefano Longhi}

\address{Dipartimento di Fisica, Politecnico di Milano and Istituto di Fotonica e Nanotecnologie del Consiglio Nazionale delle Ricerche, Piazza L. da Vinci 32, I-20133 Milano, Italy (stefano.longhi@polimi.it)\\ and
IFISC (UIB-CSIC), Instituto de Fisica Interdisciplinar y Sistemas Complejos - Palma de Mallorca, Spain}

\begin{abstract}
  In multiparticle quantum interference, bosons show rather generally the tendency to bunch together, while fermions can not. This behavior, which is rooted in the different statistics of the particles, results in a higher coincidence rate $P$ for  fermions than for bosons, i.e.  $P^{(bos)}<P^{(ferm)}$. However, in lossy systems such a general rule can be violated because bosons can avoid lossy regions. Here it is shown that, in a rather general optical system showing passive parity-time ($\mathcal{PT}$) symmetry,  at the $\mathcal{PT}$ symmetry breaking phase transition point the coincidence probabilities for bosons and fermions are equalized, while in the broken $\mathcal{PT}$ phase the reversal $P^{(bos)}>P^{(ferm)}$ is observed. Such effect is exemplified by considering the passive $\mathcal{PT}$-symmetric optical directional coupler.
\end{abstract}

 ] 

{\it Introduction.} Parity-time ($\mathcal{PT}$) symmetry, originally introduced in quantum physics to explore non-Hermitian extensions of quantum mechanics and quantum field theories \cite{r1}, has become very popular in photonics over the past few years since its first experimental demonstration in an optical directional coupler \cite{r2,r3}. A wide variety of applications based on the concept of $\mathcal{PT}$ symmetry, including laser mode control, material engineering, optical sensing and topological light transport, have been demonstrated in the last decade using integrated photonic systems, such as coupled waveguides, gratings and microcavities (see e.g. the recent reviews \cite{r4,r5,r6,r7,r8,r9,r10} and references therein). In the majority of such applications, light behaves classically. At the classic optics level, a transition from unbroken to broken $\mathcal{PT}$ symmetric phases is observed as a non-Hermitian parameter, such as the gain/loss contrast in the system, is increased. Correspondingly, the energy spectrum (i.e. propagation constants or  resonance frequencies of supermodes of the coupled waveguide/resonator system) ceases to be entirely real and complex conjugate energies emerge. The $\mathcal{PT}$ symmetry breaking point corresponds to the appearance of an exceptional point (EP), i.e. a non-Hermitian degeneracy where two (or more) eigenvalues and corresponding eigenvectors of the Hamiltonian coalesce \cite{r7,r8}. $\mathcal{PT}$ symmetric optics in the full quantum domain, where light behaves non-classically, has received little attention so far \cite{r11,r12,r13,r14,r14bis,r15,r16,r16bis,r16tris}, and previous studies mainly focused on quantum noise  near EPs \cite{r11,r12,r14bis,r16tris}. However, the full implications of $\mathcal{PT}$ symmetry in second-quantization realm remain largely unexplored. A recent experiment \cite{r16} reported on the observation of two-photon interference effects in a passive $\mathcal{PT}$-symmetric optical directional coupler, demonstrating that the Hong-Ou-Mandel dip, arising from photon bunching, quite surprisingly shifts toward  shorter distances as the loss in the system is increased. However, $\mathcal{PT}$-symmetry breaking phase transition can not be revealed looking at the dip shift.\\
In this Letter we unravel a hidden signature of $\mathcal{PT}$ symmetry breaking phase transition in second-quantization framework by considering multi-photon quantum interference in dissipative linear optical systems. The behavior of indistinguishable quantum particles is governed by their statistics, and photons can effectively show statistics tunable between bosons and fermions \cite{r17,r18,r19}. In a multiparticle quantum interference experiment, bosons show rather generally the tendency to stick together, while fermions can not. This results in a higher coincidence rate $P$ for  fermions than for bosons, i.e. 
$P^{(bos)}<P^{(ferm)}$. However, in lossy systems such a general rule can be violated. For example, bosons can display antibuching behavior in a lossy beam splitter \cite{r20,r21}. Here we show that, in a rather general passive $\mathcal{PT}$ optical system probed by an entangled photon state, at the $\mathcal{PT}$ symmetry breaking phase transition point the coincidence rates for bosons and fermions are equalized, while in the broken $\mathcal{PT}$ phase the inequality $P^{(bos)}<P^{(ferm)}$ is reversed. The effect is exemplified by considering the passive $\mathcal{PT}$-symmetric optical directional coupler as a paradigmatic model \cite{r2,r13,r16}, which is feasible for an experimental test with quantum light.\\ 
\\
\begin{figure}[htb]
\centerline{\includegraphics[width=8.4cm]{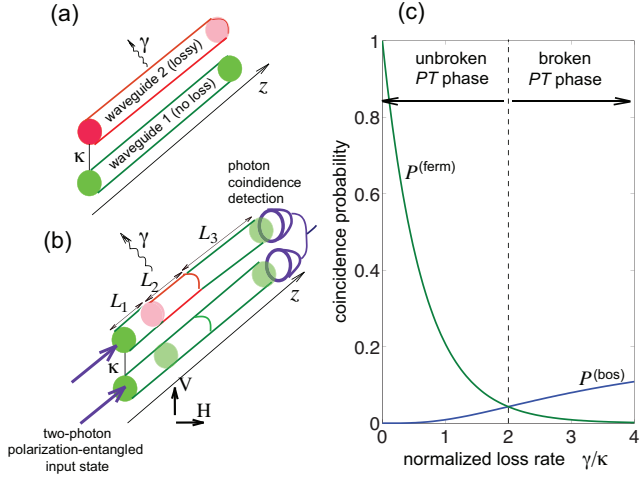}} \caption{ \small
(Color online) (a) Schematic of a passive $\mathcal{PT}$ optical directional coupler. The upper waveguide is lossy with a loss rate $\gamma$. (b) Setup for the measurement of photon coincidence and statistical signature of $\mathcal{PT}$ symmetry breaking. The first and last sections of the upper waveguide, of length $L_1= \pi/(4 \kappa)$ and $L_3= 7 \pi/(4 \kappa)$, are lossless and provide the unitary rotations $\mathcal{R}$ and $\mathcal{R}^{-1}$ of mode basis. The coupler is excited by a two-photon polarization-entangled state $(1/ \sqrt{2} ) (\hat{a}_1^{\dag (H)} \hat{a}_2^{\dag (V)} \pm \hat{a}_1^{\dag (V)} \hat{a}_2^{\dag (H)})|0 \rangle$, where $H$ and $V$ denote horizontal and vertical polarization while the $ +/-$ signs correspond to bosonic/fermionic statistics. A two-photon coincidence detection system is placed at the output ports. (c) Behavior of the coincidence probabilities $P^{(bos,ferm)}$ versus loss rate $\gamma$ for bosonic and fermionic statistics in an optical coupler with $L_2=L_1= \pi /(4 \kappa)$.}
\end{figure} 
\begin{figure}[htb]
\centerline{\includegraphics[width=8.4cm]{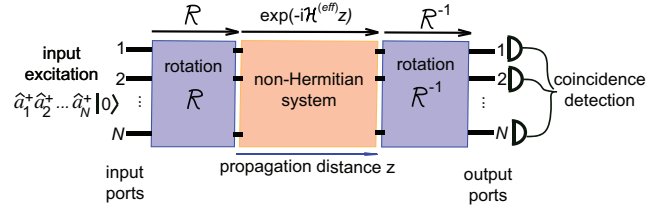}} \caption{ \small
(Color online) General setup for photon coincidence measurement. The $N \times N$-ports non-Hermitian system is excited by the $N$-particle number state $\hat{a}_1^{\dag} \hat{a}_2^{\dag}... \hat{a}_N^{\dag} |0 \rangle$. Before entering into the non-Hermitian system, the input state is rotated by the unitary operator $\mathcal{R}$. At the output ports, before detection the inverse unitary transformation $\mathcal{R}^{-1}$ is applied to the multiparticle state. The rotation $\mathcal{R}$ is defined by the Schur decomposition of the non-Hermitian matrix $\mathcal{H}^{(eff)}$ at the $\mathcal{PT}$ symmetry breaking transition point, as discussed in the text. The creation/destruction operators can effectively describe either bosonic or fermionic particles.}
\end{figure} 
{\it Non-classical light dynamics in dissipative linear optical systems.} Let us consider light dynamics is a linear passive (i.e. without gain) optical system comprising $N$ channels or nodes, such as a set of evanescently-coupled optical waveguides or resonators, and let us indicate by $\hat{a}_l$ ($\hat{a}^{\dag}_l$) the destruction (creation) operators of photons in the $l$-th channel of the system. Dissipation is described by coupling the waveguides or resonators to one or more reservoirs at zero temperature, which we assume initially in the vacuum state. To describe propagation of non-classical light in the system, one can use either the master equation or noise operator approaches (see, for instance, \cite{book}). The master equation in Lindblad form for the reduced density operator $\hat{\rho}$ of the photon field reads \cite{book,r23}
\begin{eqnarray}
\frac{d \hat{\rho}}{dz} & = &- i [\hat{H}^{(H)}, \hat{\rho}] + \sum_l \left( \hat{J}_l \hat{\rho} \hat{J}^{\dag}_l-\frac{1}{2}\hat{J}^{\dag}_l \hat{J}_l \hat{\rho}-\frac{1}{2} \hat{\rho} \hat{J}_l^{\dag} \hat{J}_l \right) \nonumber \\
& = & -i \left( \hat{H}^{(eff)} \hat{\rho}- \hat{\rho} \hat{H}^{(eff) \; \dag} \right)+ \sum_l \hat{J}_l \hat{\rho} \hat{J}_l^{\dag}
 \end{eqnarray}
where $z$ is a propagation distance (for coupled waveguide systems) or time variable (for coupled resonator systems), $\hat{H}^{(H)}$ describes the coherent Hermitian evolution of the system, $\hat{J}_l$ are the jump operators, and $\hat{H}^{(eff)} \equiv \hat{H}^{(H)}-i (1/2) \sum_l \hat{J}_l^{\dag} \hat{J}_l$ is the effective non-Hermitian Hamiltonian that describes the short-time coherent evolution of the photon field before a  quantum jump. For the sake of definiteness, in the following we will consider spatial light propagation in coupled waveguide structures, so that $z$ is the longitudinal spatial propagation distance. For a linear dissipative system, $\hat{H}^{(H)}$ is a quadratic form of $\hat{a}^{\dag}_l \hat{a}_n$ with Hermitian matrix, whereas the jump operators $\hat{J}_l$ are linear combinations of the destruction operators $\hat{a}_n$. Therefore, the most general form of the effective non-Hermitian Hamiltonian $\hat{H}^{(eff)}$ reads
\begin{equation}
\hat{H}^{(eff)} = \sum_{n,l} \mathcal{H}^{(eff)}_{n,l} \hat{a}_n^{\dag} \hat{a}_l
\end{equation}
where $\mathcal{H}^{(eff)}$ is a non-Hermitian $N \times N$ matrix with eigenvalues $\lambda_{l}$ having non-positive imaginary parts. For example, in the passive $\mathcal{PT}$-symmetric optical coupler ($N=2$), with one lossy waveguide solely [Fig.1(a)], the effective non-Hermitian Hamiltonian is given by $\hat{H}^{(eff)} = \kappa( \hat{a}^{\dag}_1 \hat{a}_2+\hat{a}_2^{\dag} \hat{a}_1)-i \gamma \hat{a}^{\dag}_2 \hat{a}_2$ corresponding to the non-Hermitian matrix 
\begin{equation}
\mathcal{H}^{(eff)}= \left(
\begin{array}{cc}
0 & \kappa \\
\kappa & -i \gamma
\end{array}
\right)
\end{equation}
where $\kappa$ is the coupling constant and $\gamma$ the loss rate.  The two eigenvalues of the matrix are $\lambda_{1,2}=-( i \gamma/2) \pm \sqrt{\kappa^2-(\gamma / 2)^2}$. For classical light excitation, light dynamics in the structure is simply described by the non-Hermitian matrix $\mathcal{H}^{(eff)}$, so that the amplitudes $a_l$ of modes in the various guides satisfy the coupled-mode equations
\begin{equation}
i \frac{da_l}{dz}= \sum_n \mathcal{H}^{(eff)}_{l,n} a_n
\end{equation}
yielding the input-output relation
\begin{equation}
a_l(z)= \sum_n \mathcal{U}_n(z) a_n(0)
\end{equation}
in terms of the propagator (scattering matrix) $\mathcal{U}(z) \equiv \exp(-i\mathcal{H}^{(eff)}z)$ of the system between input ($z=0$) and output ($z=z$) planes.
On the other hand, for non-classical states of light propagation in the system requires to solve either the master equation (1) with jump operators or the Heisenberg-Langevin equations of operators  $\hat{a}_l$, which are obtained from Eq.(4) after the replacement $a_l  \rightarrow \hat{a}_l$ and adding noise operators on the right hand side of Eq.(4) \cite{book,r14bis,r24,r25}. The two descriptions are basically equivalent \cite{book,r25}, however depending on the kind of input states (pure or mixed) and quantum correlations under investigation one of the two methods can be more feasible than the other one to address the quantum problem.
\\
\par
{\it Quantum statistical signature of  symmetry breaking.}  
Let us consider an optical  structure with passive $\mathcal{PT}$ symmetry and let us indicate by $\gamma$ a parameter (or more generally a set of parameters) that measures the loss in the system, such that at $\gamma=0$ the matrix $\mathcal{H}^{(eff)}$ is Hermitian.  A typical scenario of passive $\mathcal{PT}$ symmetry-breaking is the following one: for $\gamma< \gamma_{th}$, where $\gamma_{th}$ is a threshold value, the eigenvalues $\lambda_l$ of $\mathcal{H}^{(eff)}$ are distinct and their imaginary parts ${\rm Im}(\lambda_l)$ take the same value, i.e. all modes show the same decay rate; at $\gamma= \gamma_{th}$ two (or more) eigenvalues and corresponding eigenvectors of $\mathcal{H}^{(eff)}$ coalesce, corresponding to an EP; at $\gamma> \gamma_{th}$ there is one dominant mode with lowest decay rate. For example, for the $\mathcal{PT}$ optical coupler [Fig.1(a) and Eq.(3)], the EP arises at $\gamma_{th}=2 \kappa$. The classical signatures of the EP at the symmetry breaking point, related to the coalescence of both eigenvectors and eigenvalues of  $\mathcal{H}^{(eff)}$, are well known and have been exploited for example in sensing applications \cite{r5,r7,r8}. Here we wish to disclose a major signature of the phase transition that arises at the full quantum level, i.e. involving quantum interference effects. Namely, we excite the system with the $N$ particle number state $\hat{a}^{\dag}_1 \hat{a}^{\dag}_2 ... \hat{a}^{\dag}_N |0 \rangle$ at the input port, and detect the coincidence probability $P(\gamma)={\rm{Tr}} ( \hat{\rho} \prod_{n} \hat{a}^{\dag}_n \hat{a}_n )$ versus $\gamma$, i.e. the probability to simultaneously detect a single particle in each of the  output ports. In particular, we wish to compare the coincidence probability curves $P^{(bos)}(\gamma)$ and $P^{(ferm)}(\gamma)$ when the statistics of the photon field is switched from bosonic to (effective) fermionic. While photons are bosons, as discussed in several recent works they can effectively emulate particles with bosonic, fermionic or intermediate (anyonic) statistics by exploiting entanglement \cite{r17,r18,r19}. For example, in case $N=2$ a two-photon polarization entangled state, described by either symmetric or antisymmetric states under particle exchange, effectively emulate bosonic and fermionic particle statistics, respectively \cite{r17,r18}. In any linear optical system without dissipation, i.e. at $\gamma=0$, the following inequality always holds
\begin{equation}
P^{(bos)} \leq P^{(ferm)}
\end{equation}
with $P^{(ferm)}=1$ for conservation of particle number. Inequality (6) follows from the fact that bosons can bunch together, while fermions cannot owing to the Pauli exclusion principle. In particular, in a two-port lossless system with balanced splitting (like in a 50 \% beam splitter) one has $P^{(bos)}=0$. The equality in (6) is attained for very special optical systems that realize self- or mirror-imaging between input and output planes, and we exclude such very special cases in our discussion. Remarkably, in a dissipative system the inequality (6) can be broken, and counter-intuitive effects can arise. For example, complete anti bunching of bosons in a lossy two-port system can been observed \cite{r20,r21}. The physical reason of violation of Eq.(6) is that, since bosons can bunch together, they can propagate in the structure partially avoiding the lossy regions. Hence, as compared to fermions, bosons show a larger probability to arrive at the output plane without being absorbed in the medium.\\  
The main result of this work is that, under a suitable rotation of the photon field before and after propagation in the dissipative system, described by the unitary transformations $\mathcal{R}$ and $\mathcal{R}^{-1}$ (Fig.2), the coincidence probability curves $P^{(bos)}(\gamma)$ and $P^{(ferm)}(\gamma)$ cross exactly at the symmetry breaking point $\gamma=\gamma_{th}$, with $P^{(bos)} <P^{(ferm)}$  ($P^{(bos)}>P^{(ferm)}$) below (above) the symmetry breaking point. In other words, the symmetry breaking phase transition corresponds to violation of the inequality (6) universally valid in an Hermitian system.\\
To prove such a statement, let us consider the excitation and detection setup shown in Fig.2, where the first and last stages, described by the unitary 
transformations $\mathcal{R}$ and  $\mathcal{R}^{-1}=\mathcal{R}^{\dag}$, basically realize a rotation of basis modes. The unitary matrix $\mathcal{R}$ is chosen as follows. At $\gamma=\gamma_{th}$, let us consider the Schur decomposition of the non-Hermitian matrix $\mathcal{H}^{(eff)}$, i.e. $\mathcal{H}^{(eff)}= \mathcal{W} \mathcal{Q} \mathcal{W}^{-1}$ with $\mathcal{W}$ a unitary matrix and $\mathcal{Q}$ and upper triangular matrix having on the main diagonal the eigenvalues $\lambda_l$ of $\mathcal{H}^{(eff)}$ at $\gamma=\gamma_{th}$. 
Then we take $\mathcal{R}=\mathcal{W}$. For example, for the $\mathcal{PT}$ optical coupler of Fig.1(a) one has 
\begin{equation}
\mathcal{R}= \frac{1}{\sqrt{2}} \left(  
\begin{array}{cc}
1 & -i \\
-i & 1
\end{array}
\right).
\end{equation}
Clearly, the scattering matrix (propagator) for classical light fields of the overall system in Fig.2 is given by
\begin{equation}
 \mathcal{U}_1=\mathcal{R}^{-1}\mathcal{U}(z) \mathcal{R}=\mathcal{R}^{-1} \exp(-i \mathcal{H}^{(eff)}z) \mathcal{R}.
 \end{equation}
For construction, at the symmetry breaking point $\gamma=\gamma_{th}$ the propagator $\mathcal{U}_1$ is an upper triangular matrix, while rather generally it is not for $\gamma \neq \gamma_{th}$.
The coincidence probabilities $P^{(bos)}(\gamma)$ and $P^{(ferm)}(\gamma)$ for bosonic and fermionic particles can be computed from the classical propagator $\mathcal{U}_1$ and take the simple form 
\begin{equation}
P^{(bos)}(\gamma)= |{\rm perm} (\mathcal{U}_1(z) )|^2 \; , \; \; P^{(ferm)}(\gamma)= |{\rm det} (\mathcal{U}_1(z))|^2
\end{equation}
where ${\rm perm}$ and ${\rm det}$ denote the permanent and determinant of $\mathcal{U}_1$, respectively \cite{r19,r26}. Since at $\gamma=\gamma_{th}$ the propagator $\mathcal{U}_1$ is an upper triangular matrix, its determinant and permanent do coincide, i.e. $P^{(bos)}(\gamma_{th})=P^{(ferm)}(\gamma_{th})$. On the other hand,  for $\gamma \neq \gamma_{th}$ the propagator $\mathcal{U}_1$ is not a triangular matrix, and thus $P^{(bos)}(\gamma_{th}) \neq P^{(ferm)}(\gamma_{th})$ rather generally. Since as $\gamma \rightarrow 0$ one has $P^{(bos)} < P^{(ferm)}$ because of (6), it follows that $P^{(bos)}(\gamma)<P^{(fed)}$ for $\gamma < \gamma_{th}$. Likewise, excluding special cases where $\gamma=\gamma_{th}$ is a saddle point, one has $(dP^{(bos, ferm)}/ d\gamma)_{\gamma_{th}} \neq 0$, which implies $P^{(bos)}(\gamma)>P^{(ferm)}$ for $\gamma> \gamma_{th}$.\\
\par
{\it The passive $\mathcal{PT}$ coupler}.  To illustrate the quantum statistical signature of the $\mathcal{PT}$ symmetry breaking phase transition, let us consider the passive $\mathcal{PT}$ optical coupler [Fig.1(a)], analogous to a lossy beam splitter, excited by a polarization-entangled two-photon state in either symmetric (bosonic) or antisymmetric (fermionic) state \cite{r17,r18,r19}. The system is illustrated in Fig.1(b). The first and last sections of the coupler, of length $L_1= \pi / (4 \kappa)$ and  $L_3= 7 \pi / (4 \kappa)$ respectively, are lossless and realize the rotations $\mathcal{R}$ and $\mathcal{R}^{-1}$, respectively. The middle section of the coupler of length $z=L_2$  is dissipative, with one lossy waveguide (dissipation rate $\gamma$). The expressions of the coincidence probabilities $P^{(bos, ferm)}$ can be calculated from Eq.(9), after computation of the propagator $\mathcal{U}_1$ using Eq.(8). Alternatively, they can be calculated by solving the master equation (1). We briefly sketch here the latter approach, which is mathematically more involved  but it could be  useful to study other correlation properties under more general (mixed) state excitation of the coupler and extended to consider multi-site systems \cite{r26bis}. The master equation for the coupler reads
\begin{equation}
\frac{d \hat{\rho}}{dz}=-i [ \hat{H}^{(H)}, \hat{\rho}]+ \gamma(2 \hat{a}_{2} \hat{\rho} \hat{a}_{2}^{\dag}-\hat{a}_{2}^{\dag} \hat{a}_{2} \hat{\rho}-\hat{\rho} \hat{a}_{2}^{\dag} \hat{a}_{2})
\end{equation}
where $\hat{H}^{(H)}=\kappa(\hat{a}_{1}^{\dag} \hat{a}_{2}+\hat{a}_{2}^{\dag} \hat{a}_{1})$ and where the loss rate $\gamma$ vanishes in the first and last sections of the coupler.  The master equation can be solved after expanding the density operator $\hat{\rho}$ in the basis of Fock states $ |n_1,n_2 \rangle \equiv (1/ \sqrt{n_1 ! n_2 !}) \hat{a}_1^{\dag n_1} \hat{a}_2^{\dag n_2} |0 \rangle$, with $n_{1,2} \leq 1$ in the fermionic case. For two-particle input excitation, the Hilbert space can be limited to the Fock states with $n_1+n_2 \leq 2$, and thus comprises the four states $|0,0\rangle$, $|0,1\rangle$, $|1,0\rangle$ and $|1,1\rangle$ for fermionic particles, and the six states $|0,0\rangle$, $|0,1\rangle$, $|1,0 \rangle$, $|1,1\rangle$, $|2,0\rangle$ and $|2,0 \rangle$  for bosonic particles. After setting $\rho_{n_1,n_2; m_1,m_2} \equiv \langle n_1,n_2 | \hat{ \rho} | m_1,m_2 \rangle$ and taking into account that  $\rho_{n_1,n_2; m_1,m_2}=\rho_{m_1,m_2; n_1,n_2}^*$, for fermionic particles Eq.(10) corresponds to a system of 10 differential equations for the density matrix elements, whereas for bosonic particles one obtains a system of 21 differential equations. The coincidence probability, measured at the output ports, is given by $P^{(bos,ferm)}= \rho_{1,1;1,1}(z)$ with $z=L_1+L_2+L_3$. In case of fermionic particles, the evolution equation for the element $\rho_{1,1;1,1}$ of density operator is decoupled form all other elements, and can be readily integrated with the initial condition $\rho_{1,1;1,1}(0)=1$ yielding
\begin{equation}
P^{(ferm)}(\gamma)=\exp(- 2 \gamma L_2).
\end{equation}
On the other hand, for bosonic particles the calculation is more involved since  $\rho_{1,1;1,1}$ is coupled to other 5 elements of density operator. Namely, after setting $X_1=\rho_{1,1;1,1}$, $X_2=-i \rho_{0,2;1,1}$, $X_3=-i\rho_{2,0;1,1}$; $X_4=\rho_{2,0;2,0}$, $X_5=\rho_{2,0;0,2}$ and $X_6=\rho_{0,2;0,2}$, the following coupled equations are obtained from Eq.(10)
\begin{eqnarray}
(dX_1/dz) & = & -2 \gamma X_1+2 \sqrt{2} \kappa (X_2+X_3) \nonumber \\
(dX_2/dz) & = & -3 \gamma X_2+\sqrt{2} \kappa (X_6+X_5-X_1) \nonumber \\
(dX_3/dz) & = & -\gamma X_3+\sqrt{2} \kappa (X_4+X_5-X_1) \nonumber \\
(dX_4/dz) & = & -2 \sqrt{2} \kappa X_3 \\
(dX_5/dz) & = & -2 \gamma X_5- \sqrt{2} \kappa (X_2+X_3) \nonumber \\
(dX_6/dz) & = & -4\gamma X_6-2 \sqrt{2} \kappa X_2 \nonumber
\end{eqnarray}
which should be integrated with the initial condition $X_n(0)=\delta_{n,1}$. The coincidence probability $P^{(bos)}=X_1(z)$ is then finally computed and reads
\begin{equation}
P^{(bos)}(\gamma)= \left( \sin^2 (\omega L_2) -\cos^2 (\omega L_2) \right)^2 \exp(-2 \gamma L_2)
\end{equation}
where we have set $\omega \equiv \sqrt{\kappa^2-(\gamma/2)^2}$. An inspection of Eqs.(11) and (13) clearly shows that, for an arbitrary length $L_2$ of the dissipative waveguide section, one has $P^{(bos)}=P^{(ferm)}$ at the $\mathcal{PT}$ symmetry breaking point $\gamma=\gamma_{th}= 2 \kappa$, i.e. $\omega=0$, while $P^{(bos)}<P^{(ferm)}$ for $\gamma < \gamma_{th}$ ($\omega$ real) and  $P^{(bos)}>P^{(ferm)}$ for $\gamma > \gamma_{th}$ ($\omega$ complex). As an example, Fig.1(c) shows the behavior of  $P^{(bos)}$ and  $P^{(ferm)}$ versus $\gamma$ is an optical coupler with $L_2=L_1= \pi/(4 \kappa)$. For $\gamma=0$, the coupler with an overall length $L=L_1+L_2+L_3=(9 \pi/4 \kappa)$ behaves like a 50\% lossless beam splitter and thus $P^{(ferm)}=1$ and $P^{(bos)}=0$, corresponding to the usual scenario of perfect bunching and anti-bunching for bosonic and fermionic particles at a balanced beam splitter. As the loss rate $\gamma$ is increased, the coincidence probability for boson increases while the one for fermions decreases, until they intersect at $\gamma=\gamma_{th}$ according to the general theory presented above.\\
\par

{\it Conclusion.}  In this work we unraveled a quantum statistical signature of $\mathcal{PT}$ symmetry breaking in an arbitrary linear dissipative optical system, based on multiphoton quantum interference of symmetric (bosonic) and antisymmetric (fermionic) states. We have shown that the  coincidence probabilities $P^{(bos,ferm)}$ for bosonic and fermionic particles cross exactly at the symmetry breaking phase transition point, and that the universal inequality  $P^{(bos)} \leq P^{(ferm)}$ valid in any Hermitian system is violated in the broken $\mathcal{PT}$ phase. We exemplified such results by considering a passive $\mathcal{PT}$ optical directional coupler, where quantum interference effects have been observed in a recent experiment \cite{r16}. The present work pushes the concept of $\mathcal{PT}$ symmetry breaking into the full quantum regime highlighting the role of particle statistics. Our predictions should be feasible for an experimental observation with current integrated quantum photonic technologies \cite{r16,r17,r27,r28}.\\

\end{document}